\documentstyle[12pt]{article}
\def\al{\alpha}
\def\be{\beta}
\def\ga{\gamma}

\def\la{\lambda}

\def\ch{\chi}
\def\ps{\psi}

\def\Ga{\Gamma}

\def\cl{{\cal L}}
\def\fr#1#2{{{#1} \over {#2}}}
\def\prt{\partial}

\def\vev#1{\langle {#1}\rangle}

\def\frac#1#2{{\textstyle{{#1}\over {#2}}}}

\def\lsim{\mathrel{\rlap{\lower4pt\hbox{\hskip1pt$\sim$}}
    \raise1pt\hbox{$<$}}}
\def\gsim{\mathrel{\rlap{\lower4pt\hbox{\hskip1pt$\sim$}}
    \raise1pt\hbox{$>$}}}
\def\sqr#1#2{{\vcenter{\vbox{\hrule height.#2pt
         \hbox{\vrule width.#2pt height#1pt \kern#1pt
         \vrule width.#2pt}
         \hrule height.#2pt}}}}

\newcommand{\beq}{\begin{equation}}
\newcommand{\eeq}{\end{equation}}
\newcommand{\bea}{\begin{eqnarray}}
\newcommand{\eea}{\end{eqnarray}}
\newcommand{\rf}[1]{(\ref{#1})}
 
\renewenvironment{thebibliography}[1]
 { \rm
   \begin{list}{\arabic{enumi}.}
    {\usecounter{enumi} \setlength{\parsep}{0pt}
     \setlength{\itemsep}{3pt} \settowidth{\labelwidth}{#1.}
     \sloppy
    }}{\end{list}}
 
\begin{document}
\titlepage
 
\begin{flushright}
{DF/IST-4.96\\}
{IUHET 349\\}
{November 1996\\}
\end{flushright}
\vglue 1cm
	    
\begin{center}
{{\bf CPT VIOLATION AND BARYOGENESIS \\}
\vglue 1.0cm
{O.\ Bertolami,$^a$ Don Colladay,$^b$ 
V.\ Alan Kosteleck\'y,$^b$ and R.\ Potting$^c$\\}
\bigskip
{\it $^a$Instituto Superior T\'ecnico, 
Departamento de F\'\i sica,\\}
\medskip
{\it Av.\ Rovisco Pais 1, 1096 Lisboa Codex, Portugal\\}
\bigskip
{\it $^b$Physics Department, Indiana University,\\}
\medskip
{\it Bloomington, Indiana 47405, U.S.A.\\}
\bigskip
{\it $^c$Universidade do Algarve, U.C.E.H., \\}
\medskip
{\it Campus de Gambelas, 8000 Faro, Portugal\\}
}
\vglue 1cm
 
\end{center}
 
{\rightskip=3pc\leftskip=3pc\noindent%\tenrm
We examine the effects on baryogenesis
of spontaneous CPT violation in a string-based scenario.
Under suitable circumstances,
certain CPT-violating terms can produce 
a large baryon asymmetry at the grand-unified scale
that reduces to the observed value 
via sphaleron or other dilution mechanisms.

}
 
\vfill
\newpage
 
\baselineskip=20pt
 
{\it \noindent 1. Introduction.}
Mechanisms for generation of the baryon asymmetry 
of the Universe naturally
link cosmology with ideas from particle physics. 
Simultaneous conditions that are sufficient for baryogenesis 
are the violation of baryon number, 
the violation of C and CP symmetries, and 
the existence of nonequilibrium processes
\cite{sakharov}.
These conditions can be met 
in a grand-unified theory (GUT)
through the decay of heavy states at high energy 
\cite{yoshimura,ig,we,linde,di},
through the decay of states in 
supersymmetric or superstring-inspired models 
at somewhat lower energies 
\cite{claudson,ya,be,cl,mo}, 
or via the thermalization of the vacuum energy of 
supersymmetric states \cite{affleck}. 
The conditions can also be met in the electroweak model
through sphaleron-induced transitions between inequivalent vacua
above the electroweak phase transition 
\cite{linde,kuzmin}.
Under suitable circumstances,
such transitions can dilute baryon asymmetries
generated at higher energies 
\cite{kuzmin1}.

A mechanism is known by which certain string theories may 
spontaneously break CPT symmetry \cite{kp}.
If CPT and baryon number are violated,
a baryon asymmetry could arise 
in thermal equilibrium \cite{dolgov,cohen}. 
This mechanism for baryogenesis 
would have the advantage of being otherwise independent 
of C- and CP-violating processes,
which in a GUT are typically contrived to match 
the observed baryon asymmetry and 
are unrelated to the experimentally measured 
CP violation in the standard model.

In this work,
we investigate the consequences for baryogenesis
of certain CPT-violating terms arising
in a string-based framework.
The basic effects are determined in section 2,
while dilution mechanisms are considered in section 3. 
We show that the observed baryon asymmetry 
could be reproduced via this scenario.

\vglue 0.4cm
{\it \noindent 2. CPT Violation and Baryogenesis.}
For definiteness,
we assume the source of baryon-number violation 
is one or more processes 
mediated by heavy leptoquark bosons of mass 
$M_X$ in a GUT, 
possibly supersymmetric.
The details of this theory play no essential role
in what follows.
Baryon-number violation in the early Universe 
from the leptoquarks is assumed
to be negligible below some temperature $T_D$.
However,
we do \it not \rm take $T_D \sim M_X$
\it a priori. \rm
Instead, 
we estimate the value of $T_D$ needed 
to reproduce the observed baryon asymmetry
via CPT-violating interactions.
Verification that $T_D$ is large and of order $M_X$
therefore provides a consistency check.

We take the CPT-violating interactions to arise 
from a string-based scenario,
via couplings between Lorentz tensors $T$ 
and fermions $\ch$, $\ps$
in the low-energy four-dimensional effective lagrangian
\cite{kp}.
Suppressing Lorentz indices for simplicity,
these have the schematic form 
$\cl_I \supset \la M^{-k} 
T \cdot \overline{\ps} \Ga (i \prt)^k \ch + h.c.$,
where $\la$ is a dimensionless coupling constant,
$M$ is a large mass scale 
(presumably within roughly an order of magnitude of
the Planck mass),
$\Ga$ denotes a gamma-matrix structure, 
and $(i\prt)^k$ represents the action of
a four-derivative at order $k \ge 0$.
The CPT violation appears when appropriate components 
of $T$ acquire nonzero expectation values $\vev{T}$.

For simplicity,
we limit the scope of the present work
to the subset of these CPT-violating terms
leading directly to a 
momentum- and spin-independent energy shift 
of particles relative to antiparticles.
Terms of this type can produce effects 
in neutral-meson systems that could be observed
in laboratory experiments
\cite{kp,meson}.
These terms are diagonal in the fermion fields 
and involve expectation values $\vev{T}$ 
of only the zero components of $T$:
\beq
\cl_I \supset \fr{\la \vev{T}}{M^k} 
\overline{\ps} (\ga^0)^{k+1} (i \prt_0)^k \ps + h.c.
\quad .
\label{cptbroken}
\eeq
Since no large CPT violation is observed in nature,
the expectation $\vev{T}$ must be suppressed
in the effective theory
relative to the low-energy scale $m_l$.
The suppression factor is presumably
some (non-negative) power $l$ of 
the ratio of the low-energy scale to $M$:
$\vev{T} \sim (m_l/M)^l M$. 
Since each factor of $i\prt_0$
also acts to provide a low-energy suppression,
the condition $k+l = 2$ determines the dominant terms 
\cite{kp}.
In what follows,
we consider the various values of $k$ and $l$ in turn.

In the context of baryogenesis,
we assume each fermion $\ps$ represents 
a standard-model quark of mass $m_q$ and baryon number $1/3$.
The energy splitting between a quark and its antiquark
arising via Eq.\ \rf{cptbroken}
can be viewed as a contribution to
an effective chemical potential $\mu$
that drives the production of baryon number 
in thermal equilibrium. 

To begin, 
consider a CPT-violating coupling for a single quark field. 
The equilibrium phase-space distributions 
of quarks $q$ and antiquarks $\bar q$ at temperature $T$ are 
$f_q(\vec p)=(1+e^{(E - \mu)/T})^{-1}$ and
$f_{\bar q}(\vec p)=(1+e^{(E + \mu)/T})^{-1}$,
respectively,
where $\vec p$ is the momentum 
and $E = \sqrt{m_q^2 + p^2}$.
If $g$ is the number of internal quark degrees of freedom,
then the difference between the number densities 
of quarks and antiquarks is
\bea
n_q - n_{\bar q} & = &
{g\over(2\pi)^3}
\int d^3 p ~[f_q(\vec p)-f_{\bar q}(\vec p)]
\quad 
\nonumber\\
& = &
{g\over 2\pi^2}\int_{m_q}^\infty dE\,E\sqrt{E^2-m_q^2}
\left[{1\over1+e^{(E-\mu)/T}}-{1\over1+e^{(E+\mu)/T}}\right]
\quad .
\label{barden}
\eea
The contribution to the baryon-number asymmetry 
per comoving volume is given by 
$(n_q - n_{\bar q})/3s$,
where the entropy density $s$ of relativistic particles is 
\beq
s(T) ={2\pi^2\over45} g_s(T) T^3
\quad ,\qquad
g_s (T)=\sum_B g_B\left(T_B\over T \right)^3+
\frac 78\sum_F g_F\left(T_F\over T \right)^3
\quad .
\label{entropy}
\eeq
In this expression,
the number of degrees of freedom
of relativistic bosons $B$ and fermions $F$
forming the plasma are taken to be $g_B$ and $g_F$,
respectively.
Their component temperatures are denoted $T_B$ and $T_F$,
to allow for possible decoupled particles.
The photon and quark gases have the same temperature $T$.

Consider first the case $k=0$ with $l=2$. 
This generates via Eq.\ \rf{cptbroken}
an effective chemical potential of
$\mu \sim m_l^2/M \simeq 10^{-17} m_l$.
Substitution into Eq.\ \rf{barden} 
and use of the condition $\mu \ll T$,
which holds for any reasonable decoupling temperature $T_D$,
gives a contribution to the baryon number per comoving
volume of
\beq
\fr{n_q - n_{\bar q}}{3s} \sim
{15g\over{2\pi^4 g_s(T)}}{\mu\over T}I_0(m_q/T)
\quad ,
\label{qcontrib}
\eeq
where 
\beq
I_0(r)=\int_{r}^\infty
dx\,x\sqrt{x^2-r^2}e^x (1+e^x)^{-2}
\quad .
\eeq
The integral obeys the condition $I_0(r) < I_0(0) = \pi^2/6$.

With two spins and three colors,
$g=6$ for a given quark flavor. 
The result \rf{qcontrib} applies for each flavor.
In GUT models,
$g_s \gsim 10^2$ for $T \gsim 100$ MeV.
Disregarding possible cancellations among contributions
from different flavors,
the net baryon number per comoving volume
produced in this way with three generations of  
standard-model particles is therefore 
$ n_B/s \sim (10^{-2} \mu /T) I_0(m_q/T) 
\sim (10^{-19} m_l /T) I_0(m_q/T)$.
This is far too small to reproduce the
observed value $n_B/s \simeq 10^{-10}$.
Note that choices of $l\ge 3$ would 
produce even smaller values.
We can therefore exclude 
baryogenesis with standard-model quarks 
via $k=0$ CPT-violating couplings.

Consider next the cases with $k \ge 1$.
These have CPT-violating couplings 
of the type in Eq.\ \rf{cptbroken}
involving at least one time derivative.
In thermal equilibrium,
it is a good approximation to replace each time derivative
with a factor of the associated quark energy.
This produces energy-dependent contributions 
to the effective chemical potential,
given by
\beq
\mu \sim \left(\fr {m_l} {M}\right)^l \fr {E^k}{M^{k-1}}
\quad .
\eeq
Using Eq.\ \rf{barden},
we find that each quark generates a contribution 
to the baryon number per comoving volume of
\beq
\fr{n_q - n_{\bar q}}{3s} \sim
\fr {15 g} {4 \pi^4 g_s(T)} I_k(m_q / T)
\quad ,
\label{basym}
\eeq
where 
\beq
I_k(r) = \int_{r}^\infty dx\, x\sqrt{x^2-r^2}
{\sinh(\lambda_k x^k)\over\cosh x+\cosh(\lambda_k x^k)}
\quad 
\label{ik}
\eeq
and
\beq 
\la_k = \left(\fr {m_l} {M}\right)^l 
\left(\fr {T} {M}\right)^{k-1}
\quad .
\eeq

If $k=1$, 
the dominant contribution arises when $l=1$.
Then,
$\la_1 = m_l / M \ll 1$
and we have 
\beq 
I_1(r) \approx 
\fr {m_l} M \int_{r}^{\infty}dx\,
\fr {x^2 \sqrt{x^2 - r^2}}{1 + \cosh{x}}
\quad .
\eeq
It can be shown that $I_1 < 12m_l/M$.
This means that the contribution to
$n_B/s$ from the $k=1$ terms is again too small
to reproduce the known baryon asymmetry.

If $k \ge 2$,
the dominant contribution appears when $l=0$.
This gives $\la_k = (T/M)^{k-1}$. 
Assuming the decoupling temperature $T_D$
is well below the scale $M$, 
the integral $I_k$ has integrand peaking near $x \sim 1$
and exponentially suppressed in the region 
$1 \ll x < M/T$.
It diverges for $x > M/T$.
Physically,
different values of $x$ allow 
for contributions of fermions of different energies
$E = xT$ to the processes generating baryon number.
The divergence of the integrals for $E > M$ is 
evidently an unphysical
artifact of the low-energy approximation.
Since few particles have energy near $M$
at temperatures much less than $M$,
the integrands can safely be truncated 
above the region $T \ll E < M$.
The integrals become 
\beq
I_k(r) \approx  
\left(\fr T M\right)^{k-1}
\int_{r}^{\infty}dx\, \fr{x^{k+1}\sqrt{x^2 - r^2}}
{1 + \cosh{x}}
\quad .
\eeq
This shows that baryogenesis is more suppressed
as $k$ increases from the value $k=2$.

If $k=2$ is assumed,
then 
$\la_2 = T/M$. 
A good estimate of the value of the integral $I_2(m_q/T)$ 
can be obtained by setting $m_q/T$ to zero,
since the fermion mass either vanishes or is much smaller 
than the decoupling temperature $T_D$. 
We obtain
$ I_2(m_q/T) \approx I_2(0) \simeq 7 \pi^4 T/15 M$.
Combining this with Eq.\ \rf{basym}
produces for six quark flavors
a baryon asymmetry per comoving volume given by
\beq
\fr{n_B}{s} \sim
\fr {21 g} {2 g_s(T)} \fr {T}{M}
\simeq \fr 3 5 \fr {T}{M}
\quad .
\label{eq:eta}
\eeq
For an appropriate value of the decoupling temperature $T_D$,
it follows that the observed baryon asymmetry 
can be matched provided the interactions 
violating baryon number are still in
thermal equilibrium at this temperature.
In estimating the value of $T_D$,
the effects of dilution mechanisms
must be taken into account.
We do this in the next section.
Note that for $k\ge 3$ the extra suppression
by powers of $T/M$ further raises
the decoupling temperature $T_D$ required.

\vglue 0.4cm
{\it\noindent 3. Dilution Mechanisms.}
A potentially important source of baryon-asymmetry dilution
is the occurrence of sphaleron transitions,
which violate baryon number.
These processes are expected to be unsuppressed
at temperatures above the electroweak phase transition
\cite{kuzmin}.

Denote the total baryon- and lepton-number densities by
$B$ and $L$, respectively.
We assume that the GUT conserves the quantity $B-L$.
Sphaleron-induced baryon-asymmetry dilution
occurs when $B-L$ vanishes
\cite{kuzmin1}.
The dilution can be estimated by calculating the expectation 
of the baryon number density 
using standard model fields in thermal equilibrium 
at the temperature $T_S$
where the sphaleron transitions freeze out. 

Consider $N$ generations of 
quarks with masses $m_{q_i}$
and leptons with masses $m_{l_i}$,
$i = 1, \ldots, N$.
The free energy in a unit volume for the system 
in equilibrium at temperature $T$ is given by
\beq
{\cal{F}}= 6\sum_{i=1}^{2N}F(m_{q_i},\mu)
+\sum_{i=1}^N [2F(m_{l_i},\mu_i)+F(0,\mu_i)]
\quad ,
\label{free}
\eeq
where the parameters $\mu$ and $\mu_i$ are the chemical potentials 
of the quarks and the $i$th lepton, respectively.
Note that these are true chemical potentials here,
unlike the effective chemical potential used in the
preceding section.
In the expression \rf{free},
the free energy in a unit volume
for each constituent fermion field 
of mass $m$ and chemical potential $\mu$ is
given by the standard expression
\begin{equation}
F(m,\mu)=-T\int{d^3k\over(2\pi)^3}\left[\ln(1+e^{-(E-\mu)/T})
+(\mu\rightarrow-\mu)\right]
\quad ,
\end{equation}
where $E$ is the energy of a fermion with momentum $\vec k$.

Sphaleron transitions preserve the $N$ quantities
$L_i = l_i - N^{-1}B$,
where the individual lepton-number densities are denoted $l_i$.
In thermal equilibrium,
this leads to the relation $\mu = - \sum_i \mu_i/3N$.
Since the sphaleron freeze-out temperature $T_S$ 
is larger than any fermion mass, 
the free energy in a unit volume
can be well approximated by 
\beq
F(m,\mu) \approx F(m,0) - \fr 1 {12} 
\mu^2 T^2(1-{3\over2\pi^2}{m^2\over T^2})
\quad .
\label{freeapp}
\eeq
The conserved number densities $L_i$ are therefore given by
\beq
L_i = - \fr {\prt {\cal F}}{\prt \mu_i} 
\approx
- \fr {\mu T^2}{3N} \al + \fr {\mu_i T^2} {2} \be_i
\quad ,
\eeq
where 
\beq 
\alpha\equiv
2N - {3\over2\pi^2} \sum_{i=1}^{2N} {m_{q_i}^2\over T^2}
\quad , \qquad
\beta_i\equiv 
1-{1\over\pi^2}{m_{l_i}^2\over T^2}
\quad .
\eeq
Solving for $\mu_i$ and summing over $i$ 
leads to the expression
\beq
\mu = - \fr {6}{T^2} 
\left(\sum_{i=1}^N \fr {L_i} {\be_i}\right) 
\left( 
9N + \fr {2 \al}{N} \sum_{j =1}^N \fr 1 {\be_j}
\right)^{-1}
\quad .
\eeq
Since each quark carries baryon number 1/3,
the expectation of baryon density is
\cite{kuzmin1}
\bea
B &=& - 2 \sum_{i=1}^{2N}
\fr {\partial F(m_{q_i},\mu)} {\partial \mu} 
\nonumber \\
&=& -2 \al \left( \sum_{i=1}^N \fr {L_i} {\be_i} \right)
\left(
9N + \fr {2 \al}{N} \sum_{j =1}^N \fr 1 {\be_j}
\right)^{-1}
\nonumber \\
&\approx &\cases{\displaystyle
- {4\over 13\pi^2}\sum_{i=1}^N L_i{m_{l_i}^2\over T^2} 
\quad , \quad
& 
$B-L =0$
\quad,
\cr \displaystyle 
\frac 4 {13}(B-L) 
\quad , \quad
&
$B-L \ne0$
\quad .
\cr}
\label{dilution}
\eea
In the last step,
only the leading-order contribution has been kept.

Consider first the case where $B-L=0$ initially.
Taking the leptoquark decays 
to be dominated by the heaviest lepton of mass $m_L$
\cite{kuzmin1},
it follows from Eq.\ \rf{dilution} that
the baryon- and lepton-number densities
are diluted through sphaleron effects
by a factor of approximately $4(N-1)m_L^2/13\pi^2 NT_S^2$.
Combining this result with Eq.\ \rf{eq:eta}
produces at the present epoch
a net contribution from three generations
to the magnitude of 
the baryon-number asymmetry per comoving volume of
\beq
\fr{n_B}{s} \sim
\fr {28 g} {13 \pi^2 g_s(T_D)} \fr {m_L^2 T_D}{T_S^2M}
\quad .
\label{result}
\eeq
Taking the heaviest lepton to be the tau 
and the freeze-out temperature $T_S$
to be the electroweak scale,
this means the baryon asymmetry produced via GUT processes
is diluted by a factor of about $10^{-6}$.
Thus,
the observed value of the baryon asymmetry 
can be reproduced if,
in a GUT model conserving $B-L$ with $B-L=0$ initially,
baryogenesis takes place via $k=2$ CPT-violating terms 
at a decoupling temperature $T_D \simeq 10^{-4} M$,
followed by sphaleron dilution.
This value of $T_D$ is close to the GUT scale
and leptoquark mass $M_X$,
as is required for consistency.

Note that in obtaining Eq.\ \rf{result}
we have used the estimate of the baryon asymmetry
obtained in Eq.\ \rf{eq:eta} of section 2,
which neglects any possible effects 
from sphalerons occurring at the GUT scale.
The sphaleron transition rate at high temperatures $T$
is $\Ga \approx \al_W^4 T^4$ \cite{ambjorn},
where $\al_W$ is the electroweak coupling constant.
This implies that the rate of baryon-number violation
exceeds the expansion rate of the Universe 
\beq
H\approx \sqrt{g_s} \fr {T^2} M
\quad 
\label{H}
\eeq
for temperatures below $\al_W^4 M \simeq 10^{12}$ GeV
\cite{rs}.
Sphaleron effects at the GUT scale can therefore 
safely be disregarded.

If instead we examine the case $B-L\ne0$,
Eq.\ \rf{dilution} shows that 
essentially no sphaleron dilution occurs.
A less attractive possibility then could be countenanced: 
baryogenesis at $T_D$ as above, 
but with the asymmetry introduced via initial conditions.
In this case dilution might occur through other mechanisms,
such as the decay of the dilaton in string theories
\cite{yoshimura1,bento}.

If Eq.\ \rf{result} is to hold,
then at the GUT scale
the leptoquark interactions that violate baryon number 
must still be in thermal equilibrium with respect to 
the expansion rate of the Universe.
Suppose baryon number is violated 
via (direct and inverse) leptoquark decays
and scattering,
occurring with gauge-coupling strength $\alpha_X$.
Then,
the rates $\Ga_D$ for decay and 
$\Ga_S$ for scattering at temperature $T > M_X$
are (see, for instance, ref.\ \cite{we}):
\beq
\Ga_D\approx g_s 
\fr{\al_X M_X^2}{\sqrt{T^2 + M_X^2}}
\quad , \qquad
\Ga_S \approx g_s 
\fr{\al_X^2 T^5}{(T^2 + M_X^2)^2}
\quad .
\eeq
These rates are to be compared with 
the expansion rate $H$ of the Universe,
Eq.\ \rf{H}.
With the above decoupling temperature $T_D$
and a reasonable coupling $\alpha_X$,
both $\Ga_D/H$ and $\Ga_S/H $ exceed one
and so the decay and scattering
processes are indeed in thermal equilibrium
at the GUT scale. 

As an aside,
we remark that for $k=2$ the decoupling temperature $T_D$
is low enough for baryogenesis to be compatible with  
primordial inflationary models of the chaotic type 
and possibly also with new inflationary models.
The examples given in refs.\ \cite{goncharov,ross} 
are also consistent with COBE bounds 
on the primordial energy-density fluctuations 
and with the upper bound on the reheating temperature
that avoids the overproduction of gravitinos 
\cite{ross,bento1}.

\vglue 0.4cm
{\it\noindent 4. Summary.}
In this work,
we have explored the possibility that baryogenesis
involves spontaneous CPT breaking 
arising in a string-based framework.
In the presence of interactions that violate baryon number,
the CPT-breaking terms with $k=2$ 
appearing in Eq.\ \rf{cptbroken}
can generate a large baryon asymmetry
with the Universe in thermal equilibrium at the GUT scale.
If the interactions preserve $B-L=0$,
the subsequent sphaleron dilution
reproduces the observed value of the baryon asymmetry.

\vglue 0.4cm
We thank A. Krasnitz for discussion.
O.B.\ and R.P.\ thank Indiana University for hospitality.
This work was supported in part
by the Junta Nacional de Investiga\c c\~ao Cient\'\i fica
e Tecnol\'ogica (Portugal) 
under grant number CERN/P/FAE/1030/95
and by the U.S.\ Department of Energy 
under grant number DE-FG02-91ER40661.

\vglue 0.4cm

\end{document}